\documentclass[usenatbib]{emulateapj}
\usepackage{amssymb}
\usepackage{amsmath}
\submitted{{\it accepted by ApJ}}
\shortauthors{Ding, Yuan \& Liang} \shorttitle{Electron acceleration
in hot accretion flows}
\begin{document}

\title{Electron heating and acceleration by magnetic reconnection in hot
accretion flows}

\author{Jian Ding\altaffilmark{1}, Feng Yuan\altaffilmark{1},
and Edison Liang\altaffilmark{2}}
\altaffiltext{1}{Key Laboratory for Research in Galaxies and
Cosmology, Shanghai Astronomical Observatory, Chinese Academy of
Sciences, 80 Nandan Road, Shanghai 200030, China; fyuan@shao.ac.cn}
\altaffiltext{2}{Department of Physics and Astronomy, Rice
University, Houston, Texas, 77005, USA}


\begin{abstract}

Both analytical and numerical works show that magnetic reconnection
must occur in hot accretion flows. This process will effectively
heat and accelerate electrons. In this paper we use the numerical
hybrid simulation of magnetic reconnection plus test-electron method
to investigate the electron acceleration and heating due to magnetic
reconnection in hot accretion flows. We consider fiducial values of
density, temperature, and magnetic parameter $\beta_e$ (defined as
the ratio of the electron pressure to the magnetic pressure) of the
accretion flow as $n_{0} \sim 10^{6} {\rm cm^{-3}}$, $T_{e}^0\sim
2\times 10^9 {\rm K}$, and $\beta_e=1$. We find that electrons are
heated to a higher temperature $T_{e}=5\times 10^9$K, and a fraction
$\eta\sim 8\%$ of electrons are accelerated into a broken power-law
distribution, $dN(\gamma)\propto \gamma^{-p}$, with $p\approx 1.5$
and $4$ below and above $\sim 1$ MeV, respectively. We also
investigate the effect of varying $\beta$ and $n_0$. We find that
when $\beta_e$ is smaller or $n_0$ is larger, i.e, the magnetic
field is stronger, $T_e$, $\eta$, and $p$ all become larger.

\end{abstract}


\section{Introduction}

Magnetic reconnection serves as a highly efficient engine which
converts magnetic energy into plasma thermal and kinetic energy. The
first theoretical model for magnetic reconnection was proposed by
Sweet and Parker \citep{Parker 1958}. Ever since then, theoretical
studies along this line have been reported in various papers
including analytical theories and numerical simulations
\citep{Parker 1963, Drake 2005, Litvinenko 1996}.

Magnetic reconnection has been widely used to explain explosive
phenomena in space and laboratory plasmas such as solar flares
\citep{Somov 1997}, the heating of solar corona \citep{Cargill
1997}, and substorms in the Earth's magnetosphere \citep{Bieber
1982, Baker 2002}. Magnetic reconnection also plays an important
role in high-energy astrophysical environments such as the
magnetized loop of the Galactic center \citep{Heyvaerts 1988}, jets
in active galactic nuclei \citep{Schopper 1998, Larrabee 2003,
Lyutikov 2003}, and $\gamma$ ray bursts \citep{Michel 1994}. For
example, by assuming that particle acceleration is due to direct
current (DC) electric field and is balanced by synchrotron radiative
losses, Lyutikov (2003) found that the maximum energy of electrons
is $\sim 100$ TeV. Larrabee calculated the self-consistent current
density distribution in the regime of collisionless reconnection at
an X-type magnetic neutral point in relativistic electron-positron
plasma and studied the acceleration of non-relativistic and
relativistic electron populations in three-dimensional tearing
configurations.

But so far there is very few work on the magnetic reconnection in
hot accretion flows. In accretion flows, the gravitational energy is
converted into turbulent and magnetic energy due to the
magnetorotational instability \citep[MRI;][]{Balbus 1991}. Both
analytical works \citep[e.g.,][]{Quataert 1999, Goodman 2008,
Uzdensky 2008} and three-dimensional magnetohydrodynamic simulations
of black hole accretion flow \citep{Hawley 2002, Machida 2003,
Hirose 2004} have shown that magnetic reconnection is unavoidable
and plays an important role in converting the magnetic energy into
the thermal energy of particles. In fact, turbulent dissipation and
magnetic reconnection are believed to be the two main mechanisms of
heating and accelerating particles in hot accretion flows. In the
case of large $\beta$ ($\beta$ is defined as the ratio of the gas
pressure to the magnetic pressure), the magnetic reconnection is
even likely to the dominant mechanism of electrons heating and
acceleration \citep[][]{Quataert 1999, Bisnovatyi-Kogan 1998}. On
the observational side, flares are often observed in black hole
systems. One example is the supermassive black hole in our Galactic
center, Sgr A*, where infrared and X-ray flares occur almost every
day \citep[e.g.,][]{Baganoff 2001, Hornstein 2007, Dodds-Eden 2009}.
To explain the origin of these flares, we often require that some
electrons are transiently accelerated into a relativistic power-law
distribution by magnetic reconnection within the accretion flow
\citep[e.g.,][]{Baganoff 2001, Yuan 2003, Yuan 2004, Dodds-Eden
2009, Eckart 2009} or in the corona above the accretion flow (e.g.,
Yuan et al. 2009), and their synchrotron or
synchrotron-inverse-Compton emission will produce the flares.

Regarding the study of the electron acceleration by reconnection,
recently many works have been done on the relativistic magnetic
reconnection with the approach of Particle-in-Cell (PIC) simulation.
\cite{Zenitani 2007} presented the development of relativistic
magnetic reconnection, whose outflow speed is on the order of the
light speed. It was demonstrated that particles were strongly
accelerated in and around the reconnection region and that most of
the magnetic energy was converted into a ``non-thermal'' part of
plasma kinetic energy. Magnetic reconnection during collisionless,
stressed, X-point collapse was studied using kinetic, 2.5D, fully
relativistic PIC numerical code in \cite{Tsiklauri 2007} and they
also found high energy electrons. PIC electromagnetic relativistic
code was also used in \cite{Karlicky 2008} to study the acceleration
of electrons and positrons. They considered a model with two current
sheets and periodic boundary conditions. The electrons and positrons
are very effectively accelerated during the tearing and coalescence
processes of the reconnection. All these work investigate the
circumstance of the Sun.

In this paper we use the numerical hybrid simulation of magnetic
reconnection plus test-electron method to investigate the electron
heating and acceleration by magnetic reconnection in hot accretion
flows. We would like to note that the results should also be applied
to corona of accretion flow and jet, if the properties of the plasma
in those cases are similar to what we will investigate in the
present paper. Self-consistent electromagnetic fields are obtained
from the hybrid code, in which ions are treated as discrete
particles and electrons are treated as massless fluid. Thereafter,
test electrons are placed into the fields to study their
acceleration. The input parameters include density, temperature and
magnetic field of the accretion flow. The values of density and
temperature of astrophysical accretion flows can differ by several
orders of magnitude among various objects. Here we take the
accretion flow in Sgr A* as reference \citep[see][for details]{Yuan
2003}, but we also investigate the effect of varying parameters. For
the strength of the magnetic field, we consider mainly
$\beta_{e}=1$. Here $\beta_{e}$ is defined as the ratio of the
electron pressure to magnetic pressure, $\beta_{e}\equiv
P_{e}/P_{\rm mag}$. Note that what we usually use in hot accretion
flows is the ratio of the gas pressure and the magnetic pressure,
$\beta\equiv P_{gas}/P_{\rm mag}$. The hot accretion flow is
believed to be two-temperature, with $T_{i} \gg T_{e}$. For a
reasonable value of $T_{i}=10T_{e}$ \citep[ref.][]{Yuan 2003}, the
above $\beta_{e}$ corresponds to the usual $\beta =10$. This is the
typical value in the hot accretion flow as shown by MHD numerical
simulations \citep[e.g., ][]{Hirose 2004}. Given that the value of
$\beta$ is very inhomogenous in the accretion flow \citep{Hirose
2004}, we also consider $\beta_{e}=10, 0.1$, and $0.01$.

In Section 2, we use the numerical hybrid simulation of magnetic
reconnection to get the structure of self-consistent electric and
magnetic fields. In Section 3 we use the obtained electromagnetic
fields as the background of test electrons to investigate the
electron acceleration. Finally we summarize our results and discuss
the application in interpreting the flares in Sgr A* in Section 4.

\section{The numerical hybrid simulation of magnetic reconnection}

We use the 2.5-D (two-dimensional and three components) hybrid
simulation code of Swift \citep{Swift 1995, Swift 1996} to model the
driven steady Petschek mode magnetic reconnection, except that we
use a rectangle coordinate system. Quasi charge neutrality is
assumed, which can be written as $n_{i} \thickapprox n_{e} =n$.

The momentum equation for each ion as a discrete particle is given
by

\begin{equation}
\frac{d \textbf{v}}{dt} = (\textbf{E}+ \textbf{v} \times
\textbf{B})-\nu (\textbf{u}_{i}-\textbf{u}_{e}),
\end{equation}
where $\textbf{v}$ is ion velocity, $\textbf{E}$ is the electric
field in units of ion acceleration, $\textbf{B}$ is the magnetic
field in units of the ion gyrofrequency, and $\nu$ is the collision
frequency. A finite numerical resistivity corresponding to the
collision frequency $\nu$ is imposed in the simulation domain (X
point) to trigger the reconnection. $\textbf{u}_{e}$ and
$\textbf{u}_{i}$ are the bulk flow velocities of electrons and ions.
In particular, $\textbf{u}_{i}$ is calculated by CIC (coordinate in
cell) method from $\textbf{v}$ and the electron flow speed
$\textbf{u}_{e}$ is calculated from Ampere's
law:$$\textbf{u}_{e}=\textbf{u}_{i}-(\bigtriangledown \times
\textbf{B} / n \mu _{0} e).$$

The electric field is derived from electron momentum equation
\begin{equation}
 \textbf{E}=- \textbf{u}_{e} \times \textbf{B}-\nu
(\textbf{u}_{e}-\textbf{u}_{i})- \frac{1}{n e} \bigtriangledown
P_{e}-\frac{m_{e}}{m_{i}} \frac{d \textbf{u}_{e}}{dt}.
\end{equation}
where $P_{e}$ is obtained from the adiabatic equation $P_{e}(m_
{e}n)^{5/3}= \rm constant$, $n$ is the density, and $e$ is the
charge of electron. In equation (2), $m_{e}\ll m_{i}$, so the last
term can be omitted, and thus we can define electric field
$\textbf{E}$ with equation (2).

We use Faraday's law to update the magnetic field
\begin{equation}
\frac{\partial \textbf{B}}{\partial  t} =- \bigtriangledown \times
\textbf{E}.
\end{equation}
The velocity of ion is updated at half time step using a leapfrog
scheme with a second order accuracy. The magnetic field and the
particle positions are advanced at full time step using an explicit
leapfrog trapezoidal scheme.

Initially a current sheet separates two lobes with anti-parallel
magnetic field in the $x$-direction. The current sheet is located
along $z$=0.5$L_{z}$. The sum of thermal and magnetic pressure is
uniform across the initial current sheet. Initial conditions are:

\begin{equation} \label{eq:1}
\left\{ \begin{aligned}
         \textbf{B}_{0} &= B_{0x} \tanh(z/ \delta) \textbf{e}_{x}, \\
                n_{0} &=n_{\infty}+ \frac{n_{\infty}}{ \kappa \cosh ^{2} (z / \delta)}, \\
                    T_{e} &=n_{0}T_{0}.
                          \end{aligned} \right.
                          \end{equation}
Here $\delta$ is the half width of the initial current sheet.
$B_{0x}$, $n_{\infty}$ is the initial physical variables near the
boundaries $z$=0, $L_{z}$. The initial electron temperature $T_{0}$
is constant. $\kappa$ is the ratio of the boundary density to
current sheet density and we set $\kappa=4$.

At the boundaries $z$=0 and $L_{z}$, $B_{z}$ is set to be zero, and
there is a small inflow velocity $u_{x}$ on the two boundaries. The
inflow is continuously imposed on the boundaries throughout our
simulation. Spatially, the inflow is uniform along the boundary.

For the parameters of the black hole accretion flow, we take the
accretion flow in Sgr $A^{\ast}$, the supermassive black hole in our
Galactic center, as an example \citep{Yuan 2003}. The characteristic
density, temperature, and $\beta_e(\equiv P_{e}/P_{\rm mag})$ are
$n_{e} \sim 10^{6} {\rm cm^{-3}}, T_{e} \sim 10^{9.3} {\rm K}$, and
$\beta_{e}=1.0$, respectively. Given the inhomogeneity of the
accretion flow we also consider $\beta_{e}=10, 0.1$, and $0.01$.

In the simulation the cell size in $z$-direction is chosen to be
$\triangle z=\lambda_{0}$ ($\lambda_{0}=c/ \omega_{pi0}$ is the ion
inertial length), and the cell size in the $x$-direction is
$\triangle x=2 \triangle z$. The length of the simulation region is
$L_{x}=100 \triangle x$ and $L_{z}=100 \triangle z$. The magnetic
field $\textbf{B}$ is in units of ion gyrofrequency $\Omega_{0}$,
time is in units of $\Omega^{-1}_{0}$, and $\textbf{v}$ and
$\textbf{u}$ are both in units of $V_{A}$ ($V_{A}$ is the
characteristic Alfv\'{e}nic velocity on boundaries $z$=0 or
$L_{z}$).

With the above units, we have chosen $\delta$=0.45 and
$n_{cell}$=150 particles per cell. The spatial profile of the
resistivity imposed in the simulation corresponds to a collision
frequency $\nu=\nu_{0} {\rm exp}\{-[(x-x_{0})^{2} + (z-z_{0}) ^{2}]
/ \delta ^{2} \}$. Here $x_{0}$ and $z_{0}$ are the middle point of
simulation domain, with $x_{0}$=0.5 $L_{x}$, $z_{0}$=0.5 $L_{z}$,
and $\nu_{0}$ =0.02. The time step to advance the ion velocity is
chosen to be $ \bigtriangleup t=0.1 \Omega_{0}^{-1}$. The evolution
of the configuration of the magnetic field is shown in Figure 1.
Induced electric field $E_{y}$ (which is vertical to {\em x-z}
plane) plays an important role in electron acceleration. Spatially,
the inductive electric field appears mainly in the two pile-up
regions (on the left and right hand sides of the X-point) on the
current sheet, having the shape of two circular spots of high value.
As the reconnection goes on, the outflow together with the magnetic
field moves out from the center of the simulation box because of the
magnetic frozen-in condition. Along with it, two spots of inductive
electric field also move out slowly along the current sheet in
opposite directions. Temporally, it varies little during the
electron acceleration and can be regarded as steady. Figure 2 shows
the evolution of the induced electric field.

\begin{figure}[htb]
\includegraphics[width=84mm]{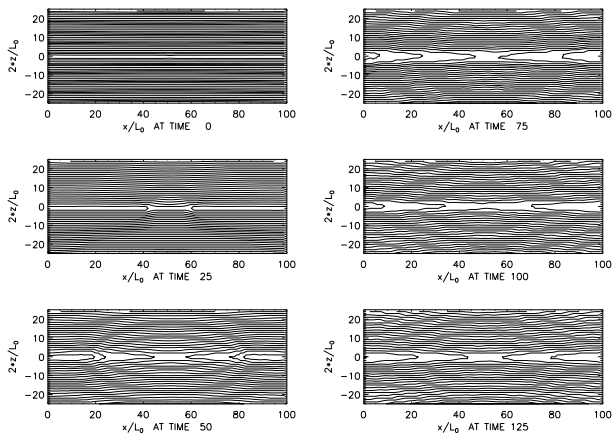}
\caption{The evolution of configurations of the magnetic field
during the reconnection process.} \label{combined lightcurves}
\end{figure}

\begin{figure}[htb]
\includegraphics[width=74mm]{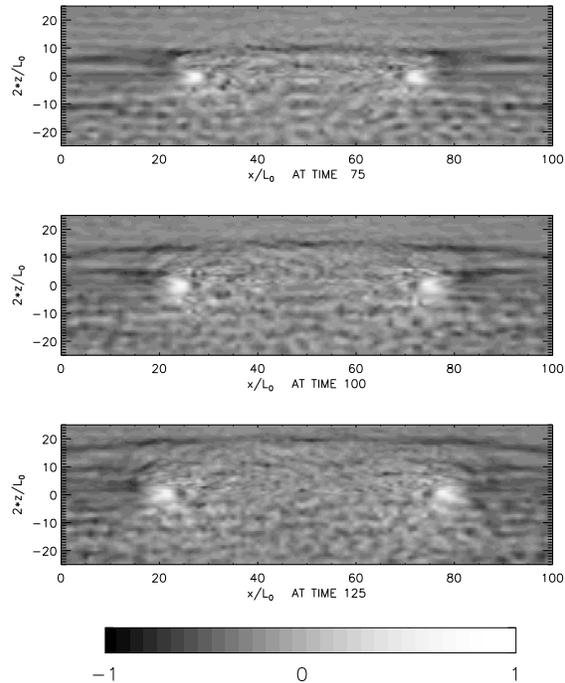}
\caption{The evolution of configuration of the inductive electric
field during the reconnection process.} \label{combined lightcurves}
\end{figure}

\section{Electrons heating and acceleration in hot accretion flows}

Magnetic reconnection is of course intrinsically a transient
process. But in our paper we treat the electron heating/acceleration
in reconnection as approximately steady. Specifically, from Figs. 1
\& 2 we see that the electromagnetic field almost remains unchanged
from t=75 to 125 and reaches a steady state. We investigate the
heating/acceleration of electrons during this interval. This
approximation is reasonable because the timescale of electron
heating/acceleration is very short compared to the timescale of the
reconnection process.

Electron dynamics in magnetic reconnection has been previously
studied using analytical theories and test particle calculations in
different magnetic- and electric-field configurations.
\citep{Litvinenko 1996, Deeg 1991, Browning 2001, Wood 2005, Birn
1994}. In such studies, the electron orbits are calculated in given
electromagnetic fields. Electrons are mainly accelerated by the
$\textbf{v} \times \textbf{B}$ reconnection electric field.
\cite{Liu 2009} performed test-particle simulations of electron
acceleration in a reconnecting magnetic field in the context of
solar flares, and found that the accelerated electrons present a
truncated power-law distribution with an exponential tail at high
energies, which is analogous to the case of diffusive shock
acceleration.

We use the test-electron method to investigate the acceleration. The
electromagnetic fields obtained from 2.5D hybrid simulation of
magnetic reconnection is used as the background. One advantage of
this method is that it saves a lot of computation time and enables
us to perform more calculations.

It is hard to study the electrons movement in the time scale $\Omega
^{-1}_{0}$ and space scale $\lambda_{0}$ in hybrid simulation. So
the cubic spline interpolation is used to get electric and magnetic
fields with higher time resolution ( 0.001 $\Omega ^{-1}_{0}$, one
step is divided into 100 steps, the step in hybrid simulation is 0.1
$\Omega ^{-1}_{0}$), suitable for the calculation for electron
dynamics. And we also use the values of electric and magnetic fields
on four grids around an electron to calculate the local electric and
magnetic fields where the electron locates. The difference method
has the following form.
$$F(x,z)=f_{11}(1- dx)(1- dz)+f_{21}(1- dx) dz$$
$$  + f_{12} dx(1- dz) + f_{22} dx dz.$$
Here the function $F$ can be any component of the electric and
magnetic fields; $f_{11,21,12,22}$ are their values on four grids
around the electron; $dx$ and $dz$ are the distance from the
electron to the nearest grid in $x$ and $z$ direction, respectively.
Such a treatment will rule out artificial high frequency
disturbance. The accuracy of such difference method is of second
order and also satisfies the requirement of our simulation. The
electron movement should satisfy the Newton equation (the
relativistic effect is considered):
\begin{equation}
\frac{d \textbf{p}}{dt} =e[ \textbf{E}+ ( \textbf{p} / \gamma)
\times \textbf{B}].
\end{equation}
Where $\textbf{p}$ is the electron momentum and $\gamma$ is the
Lorentz factor.

\begin{figure}[htb]
\includegraphics[width=84mm]{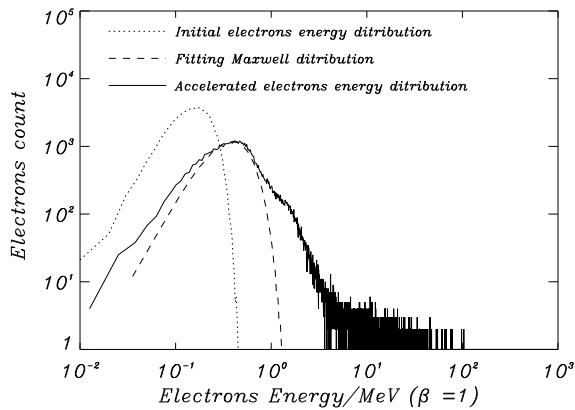}
\caption{The electrons energy distribution before and after
acceleration in the case of $\beta_{e}=1.0$. The dotted line denotes
the initial (Maxwellian) distribution. The solid line shows the
distribution after acceleration. It can be fitted by a new
Maxwellian distribution (shown by the dashed line) plus a hard
tail.} \label{combined lightcurves}
\end{figure}

The size of the simulation box is $[-50 \lambda_{0}, 50 \lambda_{0}
] \times [-25 \lambda_{0}, 25 \lambda_{0} ]$. Because of symmetry
and in order to reduce the calculation, test electrons are added is
in $[0, 50 \lambda_{0} ] \times [0, 25 \lambda_{0} ]$ the region
only. The number of the test electrons is 65536. Their initial
velocity distribution is Maxwellian and the spacial distribution is
homogeneous. The motion of test electrons is controlled by the
electric and the magnetic fields obtained in \S2.

The resulted new distribution after the acceleration in the case of
$\beta_{e}=1.0$ is shown in Figures 3 \& 4. We see that the
distribution can be described by a new Maxwellian one with higher
temperature plus a hard tail which can be fitted by a broken
power-law, i.e., the electrons are heated and accelerated after the
magnetic reconnection. The temperature of heated electrons increases
from $10^{9.30}K$ to $10^{9.65}K$. The broken power-law is described
by $dN(E) \propto E^{-p}dE$, with $p=1.47$ between $1$ MeV and $10$
MeV and $p=4$ between $10$ and 40 MeV. Most electrons are heated
rather than accelerated, and about $7.94\%$ of the total test
electrons are accelerated above 1MeV and $0.93\%$ are accelerated
above 10MeV.

\begin{figure}[htb]
\includegraphics[width=84mm]{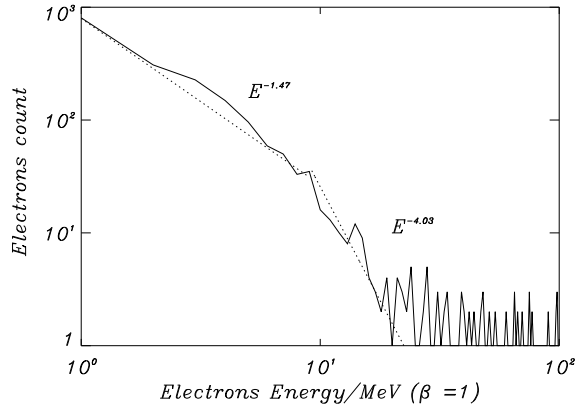}
\caption{The distribution of {\em accelerated} electrons (``hard
tail'' in Fig. 3) when $\beta=1.0$. It is fitted by a broken
power-law.} \label{combined lightcurves}
\end{figure}

\begin{figure}[htb]
\includegraphics[width=84mm]{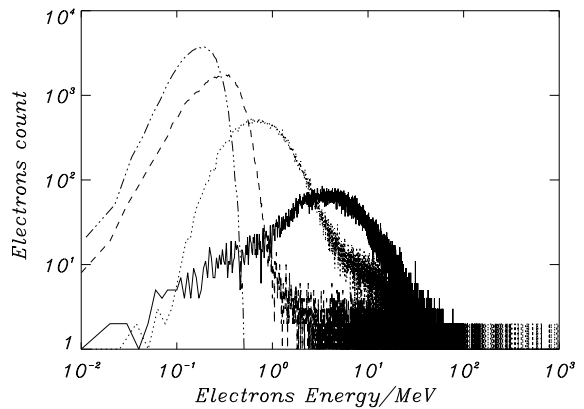}
\caption{The electrons energy distribution before
(dot-dot-dot-dashed line) and after acceleration in the cases of
$\beta =10.0$ (dashed line), $\beta=0.1$ (dotted line), and
$\beta=0.01$ (solid line). } \label{combined lightcurves}
\end{figure}

We have also calculated the cases of other values of $\beta_{e}$,
density $n_e$, and initial electron temperature $T_{e}^0$. The
results are shown in Figures 5 \& 6 and Table 1. We can see that the
results are qualitatively similar. Electrons are heated to a higher
temperature, and some of them are accelerated. The accelerated
electrons can be well described by a broken power-law distribution.
When the magnetic field becomes stronger, i.e., $\beta_{e}$ becomes
smaller, electrons are more effectively heated and accelerated,
namely the ``new'' temperature becomes higher, and more electrons
are accelerated. This implies that more magnetic energy is converted
into the kinetic energy of electrons with the increase of the
magnetic field. When $\beta$=0.1, some electrons can even be
accelerated to more than 5000 MeV. When $\beta$=0.01, however, the
magnetic field B is too strong. In this case, the accelerated
electrons reaching a certain energy will be deflected by the strong
magnetic field and run out of the acceleration region, thus further
acceleration is restrained. Although more electrons can be heated to
higher temperature and accelerated to higher energy when $\beta$
decreases from 0.1 to 0.01, the highest energy that the electrons
can reach when $\beta=0.01$ is smaller than that when $\beta$=0.1.
Another interesting result is that the ``low-energy'' part of the
broken power-law, where most accelerated electrons locate, becomes
softer with decreasing $\beta_e$, with the spectral index $p\sim
0.63, 1.47, 1.46$ and $1.91$ for $\beta_e=10.0, 1.0, 0.1,$ and
$0.01$, respectively.

\begin{table*}
\begin{center}
\label{derived} \tabcolsep 10pt \caption{Results of electrons
acceleration for different initial temperature, density and $\beta
e$} \vspace*{-0pt}
\def\temptablewidth{0.8\textwidth}
\begin{tabular*}{\temptablewidth}{@{\extracolsep{\fill}}ccccccc}
\hline \hline $T_{e}^0({\rm K})$ & $n_{e}({\rm cm}^{-3})$ &
$\beta_{e}$ &$T_{e}$(K) & power-law form & & acceleration
ratio \\
            &                   &              &           &
          & &
       \\
\hline
$10^{9.3}$&  $10^{6}$&   10&    $10^{9.55}$&    $dN(E)\propto E^{-0.63}$(1 -14 MeV) &    &$2.24\%$(E$\geq$1  MeV) \\
          &          &     &               &    $dN(E)\propto E^{-7.59}$(14-23 MeV) &    &$0.50\%$(E$\geq$10 MeV) \\
\hline
$10^{9.3}$&  $10^{6}$&  0.1&     $10^{9.85}$&    $dN(E)\propto E^{-1.46}$(1-17 MeV)&    &$44\%$(E$\geq$1 MeV) \\
          &          &     &               &    $dN(E)\propto E^{-5.05}$(17-46 MeV)&    &$3.2\%$(E$\geq$10 MeV) \\
          &          &     &               &                                      &    &$0.5\%$(E$\geq$1000 MeV) \\
\hline
$10^{9.3}$&  $10^{6}$& 0.01&    $10^{10.5}$&    $dN(E)\propto E^{-1.91}$(5-20 MeV)&    &$63.1\%$(E$\geq$5 MeV) \\
          &          &     &               &    $dN(E)\propto E^{-3.80}$(20-100 MeV)&    &$28.8\%$(E$\geq$10 MeV) \\
\hline
$10^{9.3}$&  $10^{6}$&    1&    $10^{9.65}$&    $dN(E)\propto E^{-1.47}$(1 -10 MeV)&    &$7.94\%$(E$\geq$1   MeV) \\
          &          &     &               &    $dN(E)\propto E^{-4.03}$(10-40 MeV)&    &$0.93\%$(E$\geq$10  MeV) \\
\hline
$10^{9.3}$&  $10^{7}$&    1&    $10^{9.75}$&    $dN(E)\propto E^{-1.63}$(1-10 MeV)&    &$17.3\%$(E$\geq$1 MeV) \\
          &          &     &               &    $dN(E)\propto E^{-3.25}$(10-35 MeV)&    &$1.1\%$(E$\geq$10 MeV) \\
\hline
$10^{8.5}$&  $10^{6}$&    1&    $10^{9.25}$&    $dN(E)\propto E^{-1.69}$(0.5-2.1 MeV)&    &$5.0\%$(E$\geq$0.5 MeV) \\
          &          &     &               &    $dN(E)\propto E^{-4.84}$(2.1-4.5 MeV)&    &$1.6\%$(E$\geq$1 MeV) \\
\hline \hline
\end{tabular*}
\end{center}
\end{table*}

\begin{figure}[htb]
\includegraphics[width=84mm]{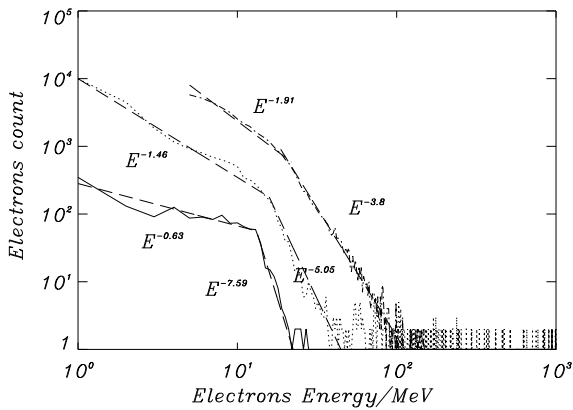}
\caption{The distribution of accelerated electrons in the cases of
$\beta=10.0$ (solid line), $\beta=0.1$ (dotted line), and
$\beta=0.01$ (dash-dotted line). All of them can be fitted by a
broken power-law (dashed line). } \label{combined lightcurves}
\end{figure}

When the density $n_{e}$ increases (but $\beta_e$ and $T_{e}^0$
remain the same), the effect is similar to the case of decreasing
$\beta_{e}$, namely the electrons are more effectively heated and
accelerated. This is because larger density implies stronger
magnetic field which is the dominant factor determining the heating
and acceleration. When the initial temperature $T_{e}^0$ decreases
(from $10^{9.3}$K to $10^{8.5}$K) but density and $\beta_{e}$ remain
unchanged, we find that the heating and acceleration become less
efficient. The ``heated'' temperature is lower, and less electrons
are accelerated. Only a fraction $\sim 5\%$ of electrons can be
accelerated to more than 0.5 MeV. This is because a lower $T_{e}^0$
corresponds to a weaker magnetic field. But surprisingly, the
low-energy part of the power-law distribution becomes softer which
seems to indicate that magnetic field is not the only factor
determining the acceleration.

\section{Summary and Discussion}

In this paper we use hybrid simulation code of magnetic reconnection
plus test-electron method to study the electrons heating and
acceleration by magnetic reconnection process in hot accretion
flows. The self-consistent electromagnetic fields are obtained from
the hybrid simulation and they are then used in the test-electron
calculation. The fiducial parameters of the background accretion
flows we adopt are density $n_{e}=10^6 {\rm cm}^{-3}$ and electron
temperature $T_{e}^0=2\times 10^{9.3}$K (ions temperature $T_{i}=10
T_{e}^0$). These values are taken from the accretion flow in Sgr A*,
the supermassive black hole in our Galactic center
\citep[ref.][]{Yuan 2003}. For the strength of the magnetic field in
the accretion flow, we set $\beta_{e} (\equiv P_{e}/P_{\rm
mag})=1.0$. This is the typical value according to the MHD numerical
simulation of hot accretion flow. We find that the electrons are
heated and accelerated by the reconnection. The new distribution can
be fitted by a Maxwellian distribution plus a broken power-law. The
temperature is increased to $10^{9.65}$K and the power-law is
described by $dN(E)\propto E^{-1.47}$ between 1 and 10 MeV and
$dN(E)\propto E^{-4}$ between 10 MeV and 40 MeV. The fraction of
electrons with energy above 1 MeV is $\sim 8\%$. Given that the
magnetic field and density in accretion flow are inhomogeneous, we
also consider $\beta_{e}=10, 0.1$, and $0.01$ and $n_{e}=10^7 {\rm
cm}^{-3}$. We find that the results are qualitatively similar,
namely the distribution of electrons can be fitted by a Maxwellian
one with a higher ``heated'' temperature plus a broken power-law.
When the magnetic field is stronger ($\beta_{e}$ is smaller or
$n_{e}$ is larger), the heating and acceleration become more
efficient. The ``heated'' temperature is higher and more electrons
are accelerated. In addition, the ``low-energy'' part of the broken
power-law becomes softer (refer to Table 1).

In our work we assume ideal MHD for the hybrid simulation, thus the
Joule heating to the accretion flow is neglected. Obviously the
electrons will be heated to a higher temperature when this effect is
considered.


\acknowledgments

We thank Drs. Pengfei Chen, Jun Lin, and especially the anonymous
referee for their useful comments and constructive suggestions. This
work was supported in part by the Natural Science Foundation of
China (grants 10773024, 10833002, 10821302, and 10825314), Bairen
Program of Chinese Academy of Sciences, and the National Basic
Research Program of China (973 Program 2009CB824800).

\end{document}